\documentclass[prd,twocolumn,showpacs,superscriptaddress,floatfix]{revtex4}

\usepackage{graphicx,psfrag,amsmath,amssymb,amsfonts,bbm,latexsym,color,dcolumn,bm}

\begin{document}

\title{Gravitational vacuum polarization phenomena due to the Higgs field}

\author{Roberto Onofrio}
\email{onofrior@gmail.com}
 \affiliation{Dipartimento di Fisica e Astronomia 'Galileo Galilei', Universit\`a di Padova, Via Marzolo 8, Padova 35131, Italy}
\affiliation{ITAMP, Harvard-Smithsonian Center for Astrophysics, 60 Garden Street, Cambridge, MA 02138, USA}

\date{\today}

\begin{abstract}
In the standard model the mass of elementary particles is considered as a 
dynamical property emerging from their interaction with the Higgs field. 
We show that this assumption implies peculiar deviations from the law of 
universal gravitation in its distance and mass dependence, as well as from 
the superposition principle. The experimental observation of the predicted 
deviations from the law of universal gravitation seems out of reach.
However, we argue that a new class of experiments aimed at studying the 
influence of surrounding masses on the gravitational force - similar to 
the ones performed by Quirino Majorana almost a century ago - could be 
performed to test the superposition principle and to give direct limits 
on the presence of nonminimal couplings between the Higgs field and the 
spacetime curvature. From the conceptual viewpoint, the violation of the 
superposition principle for gravitational forces due to the Higgs field 
creates a conflict with the notion that gravitational potentials, as assumed 
in Newtonian gravitation or in post-Newtonian parameterizations of metric 
theories, are well-defined concepts to describe gravity in their 
non-relativistic limit.

\end{abstract}

\pacs{04.25.Nx, 04.62.+v, 04.80.Cc, 14.80.Bn}

\maketitle

\section{Introduction}

The standard model of elementary particle physics has succeeded to
partially unify electromagnetic and weak interactions within a
consistent, renormalizable framework.  However, this is achieved at
the price of introducing a new interaction of all massive particles
with a scalar field, the Higgs field, which is in turn considered to
be responsible for the generation of the mass of all elementary
particles constituting matter and mediating interactions.  Such a
dynamical mechanism for generating mass should ultimately confront the
other subfield of fundamental physics in which the concept of mass
plays a crucial role, namely gravitation.  A first step towards this
confrontation has been taken in \cite{Onofrio}, in which the
implications of the Higgs field in proximity of strong sources of
gravity have been discussed in the framework of quantum field theory
in curved spacetimes \cite{Birrell}, leading to the prediction of
spectroscopic shifts in principle distinguishable from the usual
Doppler, gravitational, or cosmological shifts.
  
In this work, we continue our analysis of the Higgs interplay with
gravitation by describing, still in an elementary fashion, two
macroscopic consequences of assuming that the mass of elementary
particles is generated by the Higgs field.  Similar to \cite{Onofrio},
the basic idea is that the properties of the Higgs field in flat
spacetime are deformed by the presence of masses, and this feeds back
on the value of the masses themselves.  If the equivalence principle
is assumed to be valid, this will lead to corrections to the usual
Newton's law of universal gravitation and to violations of the
superposition principle.  In particular, in this second case we
discuss a potential relationship between such violations and already
performed experiments originally aimed at evidencing gravitational
screening.  We argue that the possible evidence for a detectable
signal from this class of experiments could be interpreted as due to
the Higgs-mediated influence of surrounding masses, at least if the
violation is large enough with respect to the one expected from the
nonlinear character of general relativity. In a more general
framework, empirical evidences for gravitational polarization of
matter have been discussed in \cite{Essen}, and our work discusses in
the standard model setting an effect related to the Higgs vacuum which
can induce gravitational vacuum polarization phenomena.

The paper is organized as follows. In Section 2, we discuss the mass
content of a body from the microscopic point of view in the framework
of the standard model of elementary particles.  It turns out that,
while most of the mass of atoms emerges from QCD vacuum, a
nonnegligible part depends instead on the vacuum expectation value
(VEV) of the Higgs field. In Section 3 this remark is used to show
that the mass of a body is affected by the mass of a surrounding body
- provided that a non-null coupling between the Higgs field and the
curvature of spacetime exists. This implies deviations from the law of
universal gravitation for the force exterted between two bodies.  In
Section 4 we discuss a more sensitive way to evidence this effect by
evaluating deviations from the superposition principle holding in
Newtonian gravity. Using the same limits as in the previous section,
we show that a window of opportunity for the experimental observation
(or to achieve better limits on the Higgs-curvature coupling) is
obtained by considering a class of experiments similar to the one
pioneered by Quirino Majorana nearly nine decades ago. While these
experiments were originally motivated by the search for gravitational
absorption/shielding, we argue that in our framework they should be
considered as the most sensitive to possible changes of inertial and
gravitational mass that a test body may experience in the presence of
nearby massive objects.  This allows us to sketch, in Section 5, two
experimental configurations in which a test mass should be maximally
coupled to surrounding masses by using a parallel plane geometry. In
the conclusions we summarize our discussions, briefly touch the
perspectives for generalizations to other extended theories of
gravity, and stress their conceptual relevance to the same definition
of gravitational potential in a Newtonian or post-Newtonian setting.

\section{A microscopic view at the mass of a macroscopic body}

In order to study the interplay between the Higgs field and
gravitation, the first question we want to answer is the following:
How is a macroscopic mass envisioned from the viewpoint of the
standard model of elementary particle physics?  At the microscopic
level, we will have elementary constituents and their binding
energies.  By neglecting the mutual gravitational and weak binding
energies among the elementary constituents, the remaining binding
energies originate from the electromagnetic interactions between
electrons and nuclei, and from the color interactions among quarks in
nucleons. In addition, there are also subleading contributions from
the residuals of the electromagnetic interaction among electrons and
nuclei of different atoms (also called {\sl van der Waals} forces) and
from the residuals of the color interactions among quarks of different
nucleons (also called {\sl nuclear} force).  Unlike quarks and
electrons, both photons and gluons are massless and are not expected
to interact with the Higgs field at tree level.  The mass of nucleons
can be then thought as being composed of a Higgs-related mass term,
and a term independent on the Higgs field.  For instance, assuming
mere additivity of the two terms, we may write the masses for protons
and neutrons
\begin{eqnarray}
m_p & = &  (2 y_u+y_d) v/(\sqrt{2}c^2)+m_{QCD}, \nonumber \\
m_n & = &  (y_u+2y_d) v/(\sqrt{2}c^2)+m_{QCD},
\end{eqnarray}
with $y_u$ and $y_d$ being the Yukawa couplings of the up and down
quarks, $v$ the VEV of the Higgs field, and $m_{QCD}$ the mass of the
nucleons due to pure gauge interaction of the gluonic fields, with
this last component independent upon the VEV of the Higgs field.  This
approximate constituent model of the nucleons can be made more
quantitative by assuming for the up and down quarks the mass values
$m_u$=2.25 MeV/c${}^2$ and $m_d$=5.0 MeV/c${}^2$, respectively. With a
VEV for the Higgs field in flat spacetime $v_0=$250 GeV we obtain, in
order to fit the proton and neutron masses, $m_{QCD}$= 928.05
MeV/c${}^2$.  This shows that, if we assume only valence quarks, their
mass is responsible for a small percentage of the nucleon inertia, of
order 1$\%$, whereas the remaining inertia is entirely due to the
gluonic field.  This model could be refined by considering the sea
quarks and partonic density functions, and in this case the gluons
will be responsible for a smaller (order of 50$\%$) percentage of
inertia, thereby sensibly increasing the percentage of inertia
sensitive to the Higgs field.  In the spirit of giving simple and
conservative estimates, in what follows we will limit our attention to
the constituent/valence model.  The mass of a body constituted of $N$
atoms, considered for simplicity homogeneously made of an element of
atomic number $Z$ and atomic mass $A$, is therefore
\begin{equation}
M=N[\gamma v/c^2+A \, m_{QCD}],
\end{equation}
where we have again distinguished between the components (electrons
and quarks) whose mass linearly depends on the Higgs VEV through a
dimensionless coefficient $\gamma=\gamma(Z,A)=1/\sqrt{2}[y_u
  (Z+A)+y_d(2A-Z)+y_eZ]$, and the ones unrelated to the Higgs field.
This mass is the one measured for instance in experiments involving
deflection of charged particles in magnetic ({\it i.e.}
non-gravitational) fields, also called {\sl inertial} mass. We will
assume that the Galileian equivalence principle holds also at the
microscopic level (see however
\cite{Lammerzahl1,Viola,Lammerzahl2,Okon} for more detailed
discussions). This hypothesis implies that also the gravitational
'charge' (also called {\sl gravitational} mass) follows exactly the
same Higgs content as its inertial counterpart, and that accurate
tests of the equivalence principle will be unable to pinpoint the
physics discussed in the next sections.

\section{Higgs field and Newton's universal gravitation law}

It is important to realize that in general the VEV of the Higgs field
depends, as for any field, on the spacetime point in which is
considered. This leads, in the presence of a specific coupling between
the Higgs field and the curvature, to a dependence of the
Higgs-dependent gravitational masses of the two bodies upon the
specific spacetime point in which they are located.  By making
explicit the Higgs-related mass contributions, the universal law of
gravitation for the amplitude of the force between two homogeneous and
spherical bodies spaced by a distance located at positions
$\vec{r}_1,\vec{r}_2$ will be written in terms of their gravitational
masses, based on Eq. 2, as\footnote{In principle one should first
  write down equations for the gravitational potential.  However,
  since the potential energy is not a useful concept if the
  superposition principle does not hold, as we will see later on, and
  since the corrections are minute, we prefer to work directly at the
  level of force corrections.}
\begin{eqnarray}
& & F_{12}(\vec{r}_1,\vec{r}_2) = G \frac{M_1 M_2}{r^2}= G \frac{N_1 N_2}{r^2} \times \nonumber \\
& &  [\gamma_1 v(\vec{r}_1)/c^2+A_1 m_{QCD}][\gamma_2 v(\vec{r}_2)/c^2+A_2 m_{QCD}],
\end{eqnarray}
where $r=|\vec{r}_2-\vec{r}_1|$. We discuss in the following possible
forms of the coupling between the Higgs field and the curvature of
spacetime. The simplest coupling is of the form ${\cal
  L}_{\mathrm{Higgs-Curvature}}=-\xi \phi^2 R$, where $R$ is the Ricci
scalar, $\phi$ the Higgs field, and $\xi$ their coupling strength
\cite{Birrell}.  Since for spherical masses the Ricci scalar in the
outer volume is zero, a Higgs-curvature coupling like $\xi \phi^2 R$
will not generate any change in the Higgs VEV, ruling out any possible
experiment or observation involving symmetrical masses.  A coupling of
this form is not experimentally observable unless specific geometries
with partial spatial symmetry are considered. However, there are
gravity models in which a scalar field may couple to the metric also
through other curvature invariants, for instance the
Einstein-Gauss-Bonnet theory which has been introduced as a promising
framework to tame ultraviolet divergences in the quantized counterpart
of the theory \cite{Lovelock}.  Experiments with higher degree of
spatial symmetry could be sensitive to the Higgs coupling to non-zero
curvature scalars such as the Kretschmann invariant
$K=R_{\mu\nu\rho\sigma}R^{\mu\nu\rho\sigma}$.  The Higgs-Kretschmann
coupling becomes expressed in terms of a single free parameter if the
Kretschmann invariant contributes to the Lagrangian through its
square-root ${\cal L}_{\mathrm{Higgs-Curvature}}=-\xi^{\prime} \phi^2
K^{1/2}$ as discussed in a strong-gravity setting (in proximity of a
back-hole) in \cite{Onofrio}.  This choice could allow us to
conjecture a minimalistic scenario, with $\xi^{\prime}$ being the only
free parameter, but is is inconsistent since such a Lagrangian term
gives rise to singularities in the $K \rightarrow 0$ limit as in a
weak-gravity scenario discussed here. A linear coupling to the
Kretschmann invariant has a consistent weak-gravity limit but requires
instead, for dimensional reasons, the presence of a further parameter
in the form of a fundamental length.  We then assume, as a natural
setting, a Lagrangian term of the form
\begin{equation}
{\cal L}_{\mathrm{Higgs-Curvature}}=-\xi_K \Lambda_\mathrm{Pl}^2 \phi^2 K,
\end{equation}
where $\Lambda_\mathrm{Pl}=(G\hbar/c^3)^{1/2}$ is the Planck length,
whose value is $\Lambda_\mathrm{Pl} \simeq 10^{-35}$ m in conventional
quantum gravity models. Of course, any length obtained by multiplying
$\Lambda_\mathrm{Pl}$ by a reasonably large number could also play its
role, or larger values of $\Lambda_\mathrm{Pl}$ such as the one
corresponding to models with {\sl early} unifications of gravity to
the other fundamental interactions \cite{Arkani}.  In the latter case
the Planck length occurs at the TeV scale via extra dimensions,
$\Lambda_\mathrm{Pl} \simeq 10^{-19}$ m. In order to keep an eye open
on this arbitrarity, in the following we will consider both these
somewhat extreme situations, by evaluating relevant quantities using
both values of $\Lambda_\mathrm{Pl}$.

With this assumption, we can mimic the reasoning in \cite{Onofrio} in
which the Higgs-curvature term adds to the mass parameter (introducing
explicitly $\hbar$ and $c$) $\mu^2 c^2/\hbar^2 \rightarrow \mu^2
c^2/\hbar^2+\xi_K \Lambda_\mathrm{Pl}^2 K$, which corresponds to
renormalize the Higgs mass term such as $\mu^2 \rightarrow
\mu^2(1+\xi_K \Lambda_\mathrm{Pl}^2 \lambda_\mu^2 K)$, with
$\lambda_\mu$ the Compton wavelength corresponding to the Higgs mass
term, $\lambda_\mu=\hbar/(\mu c) \simeq 2 \times 10^{-18}$ m, choosing
$\mu=160$ GeV.

While in flat spacetime and in the spontaneously broken phase the
Higgs field develops a VEV $v_0=(-\mu^2/\lambda)^{1/2}$, in the
presence of curved spacetime the VEV becomes
\begin{equation}
v=\sqrt{-\frac{\mu^2(1+\xi_K \Lambda_\mathrm{Pl}^2 \lambda_\mu^2 K)}{\lambda}} \simeq 
v_0\left(1+\frac{\xi_K}{2} \Lambda_\mathrm{Pl}^2 \lambda_\mu^2 K \right),
\end{equation}
the latter expression being valid in the weakly coupled regime, {\it
  i.e.} if $\xi_K \Lambda_\mathrm{Pl}^2 \lambda_\mu^2 K << 1$. Then,
by supposing that two masses are present in locations $\vec{r}_1$ and
$\vec{r}_2$, the VEV in $\vec{r}_1$ will be affected by the curvature
produced by the mass in $\vec{r}_2$ and vice versa. For the
Schwarzschild metric the Kretschmann invariant in position $\vec{r}$
due to the $j^{th}$ mass source is $K_j(r)=12 R_{s(j)}^2/r^6$,
allowing to write the VEV in location $\vec{r}_i$ as
\begin{equation}
v(\vec{r}_i) \simeq v_0 \left(1+6\xi_K \frac{\Lambda_\mathrm{Pl}^2 \lambda_\mu^2 R_{s(j)}^2}{r^6}\right),
\end{equation}
with $i,j$=1,2, $R_{s(i)}= 2G M_{i}/c^2$ being the Schwarzschild
radius of the $i^{th}$ mass.  Based on Eq. 2, this allows to write the
mass of a body as
\begin{equation}
M_i=M_i^{(0)}\left(1+6\xi_K \frac{N_i\gamma_i v_0}{M_i^{(0)}c^2}
\frac{\Lambda_\mathrm{Pl}^2 \lambda_\mu^2 R_{s(j)}^2}{r^6}\right),
\end{equation}
where $M_i^{(0)}$ is the {\sl bare} mass in the presence of flat
spacetime, {\it i.e.} in the absence of all other $j^{th}$ mass
sources. The Newton's universal law will be consequently written as
\begin{eqnarray}
F_{12}(r) = G \frac{M_1^{(0)} M_2^{(0)}}{r^2} 
& & 
\left(1+6 \xi_K \frac{N_1\gamma_1 v_0}{M_1^{(0)}c^2}\frac{\Lambda_\mathrm{Pl}^2 \lambda_\mu^2 R_{s(2)}^2}{r^6}\right)
\times \nonumber \\
& & 
\left(1+6 \xi_K \frac{N_2\gamma_2 v_0}{M_2^{(0)}c^2}\frac{\Lambda_\mathrm{Pl}^2 \lambda_\mu^2 R_{s(1)}^2}{r^6}\right).
\end{eqnarray}
Notice the cross-terms in which the two masses appear simultaneously
(since the Schwarzschild radius depends on mass), which in principle
invalidate the linear dependence of the force on the mass of each
body. Also, we will have $r^{-8}$ and $r^{-14}$ distance dependences
superimposed to the usual Newtonian $r^{-2}$ scaling.  The Higgs
correction is expected to be extremely small under practical
circumstances so that to the next-to-leading order we can keep only
the $r^{-8}$ term.  This means that the gravitational force will be
written as
\begin{eqnarray}
& & F_{12}(r) \simeq G \frac{M_1^{(0)} M_2^{(0)}}{r^2} \times \nonumber \\
& & \left[1+ 6 \xi_K \frac{v_0}{c^2} 
\left(\frac{N_1\gamma_1}{M_1^{(0)}}R_{s(2)}^2+\frac{N_2\gamma_2}{M_2^{(0)}}R_{s(1)}^2\right)
 \frac{\Lambda_\mathrm{Pl}^2 \lambda_\mu^2}{r^6}\right].
\end{eqnarray}
Assuming the same atomic composition and, for simplicity, $A=2Z$
(isoscalar nuclei), this correction is maximal for equal masses $M_1^{(0)}=M_2^{(0)}=M$, 
therefore we have 
\begin{equation}
F_{12}(r) \simeq G \frac{M^2}{r^2} \left(1+12 \xi_K \frac{\delta M}{M} 
\frac{\Lambda_\mathrm{Pl}^2 \lambda_\mu^2 R_s^2}{r^6}\right),
\end{equation}
where we have defined $\delta M/M=N \gamma v_0/(M c^2)$ as the mass
fraction of a body sensitive to the Higgs field through the Yukawa
couplings of the elementary constituents, equal to $\delta M/M \simeq
1.2 \times 10^{-2}$ for isoscalar bodies.  If we consider $M$=1 Kg
(leading to $R_s=1.5 \times 10^{-27}$ m), $\lambda_\mu=2 \times
10^{-18}$ m, and $r$=10 cm, we obtain, for the second term in Eq. 10,
the quantity $10^{-154} \xi_K$ in the usual quantum gravity scenario
with $\Lambda_\mathrm{Pl} \simeq 10^{-35}$ m, and $10^{-122} \xi_K$ if
$\Lambda_\mathrm{Pl} \simeq 10^{-19}$ m is instead considered.  If any
nonminimal coupling is active, it must give rise to tiny deviations
from Newtonian gravity, and the fact that the Newton gravitational
constant is known with a relative precision of $10^{-4}$ gives already
upper bounds of $\xi_K < 10^{150}$ and $\xi_K < 10^{118}$ for the two
choices of $\Lambda_\mathrm{Pl}$, respectively.  Lower bounds on the
detectable $\xi_K$ arise from the leading corrections to the universal
law of gravitation due to its nonlinear character and quantum gravity
effects.  These have been evaluated in \cite{Donoghue}, and the
gravitational force between two bodies is then rewritten as
\begin{eqnarray}
F_{12}(r)& = & G \frac{M_1^{(0)} M_2^{(0)}}{r^2} \times \nonumber \\
& & \left[1-\frac{2G(M_1^{(0)}+M_2^{(0)})}{c^2r}-\frac{127}{10\pi^2} \frac{\hbar G}{c^3 r^2} \right].
\end{eqnarray}
Plugging in numbers for our case it turns out that the nonlinear term
(the second contribution in the square bracket of the above equation)
dominates over the quantum gravity correction (the third term linear
in $\hbar$) and represents a correction of $\simeq 3 \times 10^{-26}$
with respect to unity, forbidding the observation of effects due to
$\xi_K< 2 \times 10^{128}$ for $\Lambda_\mathrm{Pl} \simeq 10^{-35}$
($\xi_K< 2 \times 10^{96}$ for $\Lambda_\mathrm{Pl} \simeq 10^{-19}$).

The upper bounds of $\xi_K$ just discussed can be compared to the ones
obtainable in dedicated astrophysical surveys near black holes. By
repeating the analysis, carried out in \cite{Onofrio} for the case of
a square-root Kretschmann coupling to the Higgs field, relative
frequency shifts in the electronic transitions due to a linear
coupling to the Kretschmann invariant are estimated to be
\begin{equation}
\frac{\delta \nu}{\nu}= \frac{\delta m_e}{m_e} \simeq \frac{\xi_K}{2} \Lambda_\mathrm{Pl}^2 \lambda_\mu^2 K,
\end{equation}
with $m_e$ the electron mass. For spectroscopic shifts arising from
atoms in the innermost stable orbit of a black hole ($r=3R_s$), this
yields $\delta \nu/\nu \simeq 2 \xi_K \Lambda_\mathrm{Pl}^2
\lambda_\mu^2/(243 R_s^4)$.  If astrophysical surveys will rule out
the presence of anomalous frequency shifts at the $\delta \nu/\nu
\simeq 10^{-5}$ level, {\it i.e.} the current instrumental sensitivity
limit on ammonia surveys \cite{Henkel,Wilson}, one will get, for
$\Lambda_\mathrm{Pl}=10^{-35}$ m, the bound $\xi_K < 2.5 \times
10^{118}$ from the analysis of solar black holes, and $\xi_K < 1.5
\times 10^{39}$ from the analysis of mini black holes with a mass of
$10^{11}$ kg, still large enough to survive quantum evaporation up to
now (with $\xi_K < 2.5 \times 10^{86}$ and $\xi_K < 1.5 \times 10^{7}$
if $\Lambda_\mathrm{Pl}=10^{-19}$m is instead assumed).

\section{Higgs field and the superposition principle for gravitation}

A more sensitive framework to constrain possible nonminimal couplings
related to the Higgs field is provided by experiments in which the
superposition principle for Newtonian forces is tested.  This allows
for the usually more controllable and reproducible environment
characteristic of tabletop experiments. However, there are intrinsic
limitations since the superposition principle for gravitational forces
is approximately valid only within the linearized, weak-gravity limit,
but it is expected to fail if a sufficient degree of precision is
achieved, due to the underlying nonlinear character of general
relativity. As we will discuss below, this will translate into a lower
bound on the minimum detectable value of $\xi_K$.
 
Let us consider two bodies 1 and 2 separated by a distance $r$,
gravitationally attracting each other in the presence of a third body
3.  This last is put in proximity of body 2, at a distance $R$, such
that the line passing along 2 and 3 is orthogonal to the line passing
along 1 and 2.  In this way the force component along the direction
joining bodies 1 and 2 is not affected by the force exerted between
bodies 2 and 3. The Higgs vacuum polarization will induce a change of
the masses, and if we focus on body 2, the force exerted on it due to
body 1 will change by an amount\footnote{Obviously in our setting body
3 will also induce a mass change in body 1, and this last will also
induce changes in the mass of body 2 and 3. For simplicity, and
since we are interested to orders of magnitude estimates, we
truncate these nonlinear couplings by focusing on the mass changes
induced by body 2, our 'test' mass, by the introduction of body 3.}
\begin{eqnarray}
& & \delta F_{12} =6 \xi_K \frac{G M_1^{(0)}}{r^2} 
\left(1+6 \xi_K \frac{N_1\gamma_1 v_0}{M_1^{(0)}c^2} \frac{\Lambda_\mathrm{Pl}^2 \lambda_\mu^2 R_{s(2)}^2}{r^6}\right)
\times \nonumber \\
& & \left(1+6 \xi_K \frac{N_2\gamma_2 v_0}{M_2^{(0)}c^2} \frac{\Lambda_\mathrm{Pl}^2 \lambda_\mu^2 R_{s(1)}^2}{r^6}\right)
\frac{N_2 \gamma_2 v_0}{c^2} \frac{\Lambda_\mathrm{Pl}^2 \lambda_\mu^2 R_{s(3)}^2}{R^6},
\end{eqnarray} 
The relative force change is therefore, in the weak limit
\begin{equation}
\frac{\delta F_{12}}{F_{12}} \simeq 6 \xi_K  
 \frac{N_2 \gamma_2 v_0}{M_2^{(0)}c^2} \frac{\Lambda_\mathrm{Pl}^2 \lambda_\mu^2 R_{s(3)}^2}{R^6}.
\end{equation}
This implies a violation of the superposition principle linear in the
coefficient $\xi_K$, independent upon the distance between the two
bodies 1 and 2 and, as before, depending on the percentage of the
Higgs-related mass in a nucleon.  By considering again an element with
$A=2Z$, and a mass $M_3$=1 Kg at a distance of 10 cm, we obtain
$\delta F_{12}/F_{12} \simeq 6 \times 10^{-155} \xi_K$.  Even in this
case the expected shift is minute unless large values of $\xi_K$ are
assumed.

This has to be confronted with the deviation from the superposition
principle expected in post-Newtonian parameterizations of general
relativity. In the configuration discussed above, this becomes
\cite{Braginsky}
\begin{equation}
F_{12}=G \frac{M_1 M_2}{r^2}\left(1+\frac{R_{s(3)}}{R}\right).
\end{equation}
Therefore, we have genuine general relativity corrections directly
proportional to $R_{s(3)}/R$.  In the same example discussed above,
this leads to force shifts $\delta F/F|_{GR} \simeq 10^{-26}$, and
then values of $\xi_K \leq 10^{128}$ cannot be discriminated from this
background due to general relativity corrections.

Among precision experiments able to further constraint nonminimal
coupling coefficients, it is worth to point out that Quirino Majorana
nine decades ago \cite{Majorana} (see also \cite{Austin} for former
attempts) provided tests of the superposition principle for
gravitational forces, with a dynamical version of the experiment being
subsequently performed at about the same level of precision
\cite{Braginsky}.  More specifically, in Majorana's experiments a
reduction of the weight of a spherical mass symmetrically surrounded
by another mass in close proximity was observed at the $10^{-9}$
level.  In a first series of experiments using a 1.274 Kg mass of lead
surrounded by 115 Kg of mercury, Majorana observed a decrease in
weight of $9.7 \times 10^{-10}$ Kg \cite{Majorana1} corresponding to a
relative decrease of $7.7 \times 10^{-10}$. In a second version of the
experiment he surrounded the same lead mass with 9,603 Kg of lead,
obtaining a decrease in weight of $2.01 \times 10^{-9}$ Kg
\cite{Majorana2}, translating into a relative decrease of $1.578
\times 10^{-9}$. Majorana interpreted these results in the framework
of gravitational screening, obtaining in the first class of
experiments with mercury a gravitational permeability coefficient
about twice bigger than the one obtained in the second class.
Objections to this interpretation soon arose, with current stronger
limits coming from the study of solar eclipses and tides
\cite{Russell,Wang,Unnikrishnan} (see also \cite{Weber} for a
discussion of shielding of dynamical gravitational fields such as
gravitational waves). An alternative view in which the observed effect
is due to a mass reduction coming from proximity masses was suggested
in Ref. \cite{Russell}.  Incidentally, in this view the larger
effectiveness, relatively to the involved masses, in decreasing the
weight of the lead mass by the mercury mass with respect to the more
massive lead mass could be tentatively justified due to the closeness
of the former mass to the lead sphere, as it seems corroborated by a
detailed scrutiny of the experimental setup drawings reported in
Majorana's papers \cite{Majorana1,Majorana2}.  In our setting, by
rejecting the hypothesis that mass shifts were effectively observed,
and claiming that the superposition principle is valid at the
$10^{-9}$ accuracy level implies, based on Eq. 13, an upper bound on
the Kretschmann coupling $\xi_K < 2 \times 10^{145}$.

\begin{table}

\begin{tabular}{lll}
\hline\noalign{\smallskip}
 & $\Lambda_\mathrm{Pl}=10^{-35}$ m & $\Lambda_\mathrm{Pl}=10^{-19}$ m      \\
\noalign{\smallskip}\hline\noalign{\smallskip}
(a) 2-body gravitational force                & $10^{150}$             & $10^{118}$           \\
(b) 3-body superposition test            & $2 \times 10^{145}$    & $2 \times 10^{113}$  \\
(c) Higgs-shifts 1 M$\odot$ BH         & $2.5 \times 10^{118}$  & $2.5 \times 10^{86}$ \\
(d) Higgs-shifts mini-BH               & $1.5 \times 10^{39}$   & $1.5 \times 10^{7}$  \\
\noalign{\smallskip}\hline
\end{tabular}
\caption{Summary of the possible upper bounds on Kretschmann-Higgs
  couplings $\xi_K$ from tabletop experiments (first two rows) and
  astrophysical surveys (second two rows).  We assume current
  instrumental sensitivities equivalent to a gravitational force
  measurement measured at $10^{-4}$ precision level in (a) and
  superposition principle tested at $10^{-9}$ precision level in (b),
  both with test masses of 1 Kg at distances of 10 cm and made of an
  isoscalar element. For the cases of (c) and (d) we assume a minimum
  detectable relative frequency shifts of $\delta \nu/\nu
  =10^{-4}$. In (d) we assume the optimal situation of the lightest
  miniblack hole of mass $M=10^{11}$ kg. Upper bounds have been
  evaluated, as discussed in the text, for two different values of the
  Planck length, the standard one and one assuming early unification
  of gravity at the Fermi scale.}
\label{tab:1}       
\end{table}

More recent experiments such as the one described in
\cite{Schurr,Nolting}, while yielding stronger limits on gravitational
absorption than the Majorana's ones as discussed in
\cite{Unnikrishnan1}, do not necessarily correspond to stronger
constraints on the influence of surrounding masses as the test mass
and the surrounding bodies were significantly more spaced apart than
in Majorana's experiments\footnote{For instance, the experiments
  reported in \cite{Schurr,Nolting} utilize gold-plated copper
  cylindrical masses with diameter of 45 mm, while the inner diameter
  of the stainless steel tanks within which they are moved vertically
  is 100 mm. The tank-test mass distance is therefore larger than the
  distance between the mercury container and the copper mass in the
  first Majorana's experiment set up in Turin.}, with a similar
relative precision for mass measurements.  Experiments using rotating
torsional balances as in \cite{Schlamminger} designed for test of the
equivalence principle would be extremely sensitive to possible
differences induced by the Higgs field between inertial and
gravitational mass, assumed in this paper to be rigorously
proportional to each other. However, they are not sensitive enough to
look for the influence of surrounding bodies on the gravitational
force, as shown by the fact that the upper limit on other than purely
gravitational force for an Yukawa interaction range of order of meters
is of the order of $\leq 10^{-5}$ with respect to Newtonial gravity.

Finally, we summarize in table 1 the discussion on the four possible methods to provide 
upper bounds to the parameter $\xi_K$. Astrophysical surveys in principle provide a 
better leverage towards limiting this parameter, but we believe that improvements by 
some orders of magnitude could be achieved with dedicated tabletop experiments, as we 
will briefly suggest in the following section.

\section{Possible upgrades for high precision tests of the superposition principle}

Dedicated configurations in which violations to the superposition
principle for gravitational forces could be evidenced or ruled out at
the highest level of precision may be designed.  Since we expect the
sought effect to be proportional to the surface-to-mass ratio, the
basic idea behind these optimized configurations is to take into
account the geometrical shape of a body, and use slabs instead of the
usual spherical geometry.  Two different configurations can be
envisaged to find out or give upper bounds to a possible Kretschmann
coupling between the Higgs field and gravitation. One consists in
using a Cavendish balance with test masses designed as plates, for
instance of a rectangular shape rather than the usual spherical shape
(Fig. 1a). Two plates are also added on each side of the test mass in
a parallel plate configuration, at approximately the same distance to
compensate for the added gravitational force. The requirements on the
distance are not stringent since the gravitational force between
parallel plates is independent on distance, apart from border effects.
In this configuration the plates lie along the vertical axis, in order
to cancel any contribution due to the Earth's gravitational
field. This closely resembles the scheme proposed in \cite{Lambrecht}
for the study of submillimeter short-range non-Newtonian components of
the gravitational force.

\begin{figure}[t]
\includegraphics[width=0.9\columnwidth, clip=true]{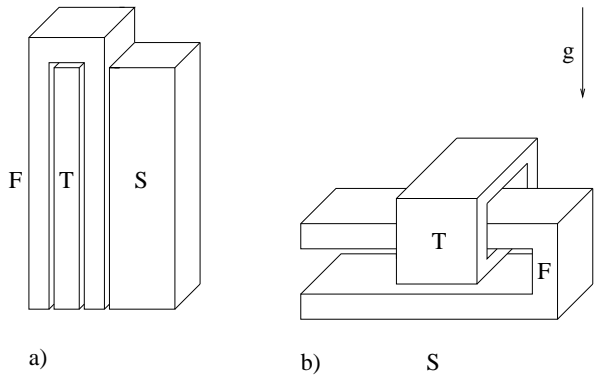}
\caption{Schematics of the active arm of possible experimental setups
  aimed at observing Higgs-related deviations from the superposition
  principle for gravitational forces.  (a) Cavendish-like balance with
  the test mass (T), the source mass (S), and a removable fork-shaped
  mass (F), with masses symmetrically centered around the test mass.
  (b) Majorana-like balance with an horizontal geometry, in which the
  source mass is the Earth.  The direction of the Earth's
  gravitational field is also evidenced.  The fork-shaped mass may be
  replaced with a variable-density rotating chopper in order to have a
  time-modulation of the effect, similarly to the scheme used in
  \cite{Braginsky}.}
\label{higgs1.fig1}
\end{figure}

The second scheme, more in the spirit of Majorana's experiments, could
alternatively exploit the Earth as a source of the gravitational
field, with a highly sensitive balance and test masses still of planar
shape but now lying along the horizontal plane (Fig. 1b).  In both
cases, the force between the gravitational {\sl source} and the {\sl
  test} mass should be monitored in the absence ({\it i.e.}, at
distance large enough) and in the presence of the two parallel plates
inserted on the opposite sides of the test mass plate.  Also, spurious
signals due to the deviations of the gravitational field due to the
finite size of the plates (analogous to the deviations of the electric
fields for finite size configurations studied in \cite{Wei}) and from
electric forces due to patch effects (in the case of conducting
surfaces, see in particular \cite{PollackKim}) or localized charges
(in the case of insulating surfaces) must be carefully modeled before
claiming a positive signal due to the Kretschmann coupling of the
Higgs field to the metric (see also \cite{OnofrioIJMPA} for a recent
general discussion of systematic effects in Casimir force
experiments).

\section{Conclusions}

In conclusion, we have discussed gravitational vacuum polarization
effects induced by the Higgs field, which imply distinctive changes in
Newton's gravitational law and the superposition principle for
gravitational forces.  Even considering the current precision level in
the determination of the universal gravitational constant $G$
\cite{Gillies}, a direct measurement of deviations from Newton's law
is experimentally precluded. Nevertheless, we believe that experiments
aimed at evidencing deviations from the superposition principle using
modern balance techniques are feasible and could shed light on
possible nonminimal couplings between the Higgs (or a Higgs-like
field) and spacetime, complementary or in alternative to dedicated
astrophysical surveys as suggested in \cite{Onofrio}.  We have also
discussed explicit experimental configurations which seem the most
promising to evidence the presence of gravitational effects due to
proximity masses.

The phenomenological platform has been developed assuming a linear
coupling of the Higgs field to the Kretschmann invariant, however
other forms of coupling could be considered, as for instance the
scalar of $3-$curvature ${}^{(3)}R$ or the trace of the extrinsic
curvature \cite{Nacir} in the framework of the recently proposed
Ho\v{r}ava-Lifshifts gravity \cite{Horava}.  An even more general
scenario left open in this analysis is the one in which the Higgs
field couples also to invariants related to torsion. For
$f(R)$-gravity models, this has intriguing implications as it
naturally leads to weak-like, chiral interactions acting at an energy
scale close or coincident with the Fermi scale
\cite{Fabbri1,Fabbri2,Capozziello}, paving the road to a possible
unification of gravitation and weak interactions.  While a
comprehensive analysis of phenomenological bounds on curvature-torsion
models is beyond the scope of this paper, this shows again that our
discussion relative to a specific, model-dependent Higgs-curvature
coupling through the Kretschmann invariant provides an operative
setting in which extensions of general relativity may confront
possible experiments/observations.  Although partially related to our
consideration, we mention recent work discussing the possible mass
shifts of heavy elementary particles due to a Higgs self-polarization
phenomenon, supposed to be observable near pair production threshold
at high energy accelerators, most notably for the top quark
\cite{Reucroft}.

Any possible matching between quantum vacuum and gravitation has been
so far frustrated by the difficulty for successful experimental
observation of predicted effects, as in the well-known example of
Hawking radiation, due to the combined smallness of the universal
constant of gravitation and Planck's constant, and the discussion in
this work confirms this state of affairs in terms of experimental
observability of the investigated effects.  It is at least comforting
to point out that, at a more conceptual level, our findings also show
a conflict between Higgs field physics and the notion that
gravitational potentials, as assumed in Newtonian gravitation or in
post-Newtonian parameterizations of metric theories, are meaningful
concepts to describe gravity in their non-relativistic limit,
providing a further source of concern for any future hypothetical
unification between quantum physics and gravitation.

\acknowledgements
We are grateful to Daniela Armocida from the Accademia Nazionale dei Lincei 
for crucial help in retrieving Refs. \cite{Majorana1,Majorana2}.

\end{document}